\documentclass[preprint,12pt]{elsarticle}

\usepackage{amssymb}

\begin{document}

\begin{frontmatter}



\title{Complexity of Planar Embeddability of Trees inside Simple Polygons}


\author{Alireza Bagheri}
\author{Mohammadreza Razzazi}

\address{Department of Computer Engineering \& IT,\\Amirkabir University of Technology, Tehran, Iran.}

\begin{abstract}
Geometric embedding of graphs in a point set in the plane is a well known problem. 
In this paper, the complexity of a variant of this problem, where the point set 
is bounded by a simple polygon, is considered. Given a point set in the plane bounded by a simple polygon 
and a free tree, we show that deciding whether there is a planar straight-line embedding of the 
tree on the point set inside the simple polygon is NP-complete. This implies that \textit{the straight-line constrained 
point-set embedding of trees} is also NP-complete, which was posed as an open problem in \cite{DDLMW:PETPD}.
\end{abstract}


\end{frontmatter}




\section{Introduction}
\label{IntroSect}
The problem of deciding whether a certain combinatorial structure can be embedded in the plane, 
as well as computing an embedding of that structure has been a recurrent theme in many fields but 
particularly in graph drawing. From a graph drawing perspective ( see \cite{BETT:ADGAB} for a survey of 
graph drawing), the traditional questions ask whether a graph can be embedded in the plane 
such that some criterion is satisfied, e.g. that the area of the resulting embedding, the number of edge 
crossings or the number of edge bends is minimized. The problem that we investigate has a slightly different 
perspective, a point set and a graph are given and we are interested in determining if the graph can be 
straight-line planar embedded onto the point set. We say that an $n$-node graph $G=(V,E)$ can be 
\textit{straight-line planar embedded} onto a set of $n$ points $P$, if there exists a one-to-one mapping 
$\phi : V \rightarrow P$ from the nodes of $G$ to the points of $P$ such that edges of $G$ intersect only at nodes. 
That is, $(\phi(u_1), \phi(v_1)) \cap (\phi(u_2), \phi(v_2)) = \emptyset$, for all $(u_1, v_1) \neq (u_2, v_2) \in E$.

Variations of this embedding problem have been investigated in the literature. The problem of characterizing 
what class of graphs can be
embedded into any point set in general position (no three points being collinear) was studied 
in \cite{GMPP:EPTVSP}. They showed that the answer is the class of outerplanar graphs, that is, graphs
that admit a straight-line planar embedding with all vertices in the outerface. This result
was rediscovered in \cite{CU:SEPGPS}, and efficient algorithms for constructing such an embedding for a
given graph and a given point set are described in \cite{B:OEOPG}. 

A tree is a special case of outerplanar graph. In this case, one also can choose to which
point the root should be mapped \cite{BMS:OAET,IPTT:RTEPPP,PT:LRT}. Bipartite embeddings of trees were 
considered in \cite{AGHNR:BETP}. If we allow each edge to be represented by a polygonal path with at most 
two bends, then it is always possible to get a planar embedding of a planar graph that maps the vertices to a
fixed point set \cite{KW:EVPFBSPG}. If a bijection between the vertices and the point set is fixed, then we need
$O(n^2)$ bends in total to get a planar embedding of the graph, which is asymptotically
tight in the worst case \cite{PW:EPGFVL,BDL:DCGOCP}. 

It is shown in \cite{C:PEVGUFPSNPH}, that deciding whether there is 
a planar straight-line embedding of a given planar graph such that its vertices are embedded onto a given
point set in the plane is NP-complete. In this paper, we prove that deciding whether there is 
a straight-line planar embedding of a given free tree onto a given point set in the plane such that 
the edges of the tree do not intersect the boundary of a given simple polygon that contains the points 
is NP-complete. This implies that the straight-line constrained 
point-set embedding of trees is also NP-complete, which was posed as an open problem in \cite{DDLMW:PETPD}. 



\newtheorem{theorem}{Theorem}
\begin{theorem}
\label{Theorem1}
Let $P$ be a set of $n$ points in the plane inside a simple polygon $S$, and let $T$ be a $n$-node free 
tree. Deciding if there exists a straight-line planar embedding of $T$ onto point set $P$ such that 
edges of $T$ do not intersect the boundary of polygon $S$ is an NP-complete.
\end{theorem}

\section{NP-completeness Proof}
\label{ProofSect}
It is clear that the problem belongs to NP. Given a straight-line planar embedding of a free tree $T$ 
onto a point set $P$ we can test in polynomial time, in the number of nodes of $T$ and the number of 
vertices of $S$, whether the embedding actually is planar and does not intersect the boundary of $S$. 
For showing the NP-hardness, the reduction is from $3$-partition.

\begin{itemize}
\item[]{\bf Problem:} $3$-partition
\item[]{\bf Input:} A natural number $B$, and $3n$ natural numbers $a_1$, ..., $a_{3n}$, 
where $B/4 < a_i < B/2$, $i=1,...,3n$.
\item[]{\bf Output:} $n$ disjoint sets $S_1$, ..., $S_n$ $\subset$ $\{a_1, ..., a_{3n}\}$, 
where $|S_j| = 3$ and $\sum_{a \in S_j} a = B$ for all $S_j$, $j=1,...,n$.
\end{itemize}

\begin{figure}
\setlength{\unitlength}{1.00mm}
\centering
\vspace{-4.5cm}
\hspace*{-3cm}
\includegraphics{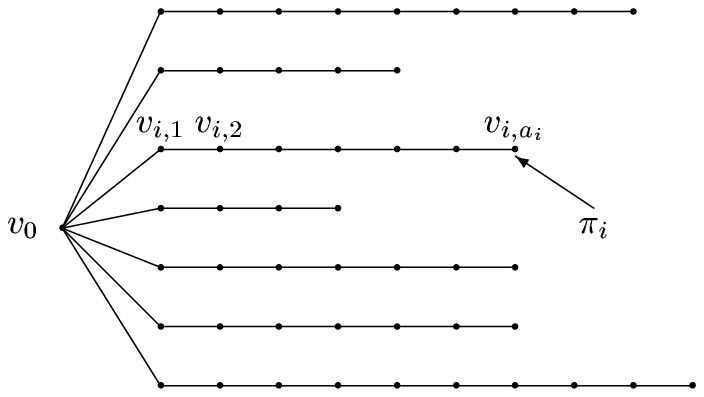}
\vspace{-21cm}
\caption{Free tree $T$ for the NP-hardness reduction.}
\label{FTreeFig}
\end{figure}

We will use that $3$-partition is a strongly NP-hard problem, that is, it is NP-hard even if $B$
is bounded by a polynomial in $n$ \cite{GJ:CI}. Observe that because $B/4 < a_i < B/2$, it does not 
make sense to have sets $S_j$ with fewer or more than $3$ elements. That is, it is equivalent to ask for
subdividing all the numbers into disjoint sets that sum to $B$. Of course, we can assume that
$\sum_{i=1}^{3n}a_i = Bn$, as otherwise it is impossible that a solution exists.
Given a $3$-partition instance, we construct the following free tree $T$ (see Figure \ref{FTreeFig}):

\begin{itemize}
\item{For each natural number $a_i$ in the input, make a path $\pi_i$ consisting of $a_i$ vertices $v_{i,1}$, ..., $v_{i,a_i}$.}
\item{Consider an additional vertex $v_0$ for $T$. For each path $\pi_i$, connect vertex $v_{i,1}$ to vertex $v_0$.}
\end{itemize}

The idea is to design a point set $P$, inside a simple polygon $S$, such that vertex $v_0$ can be mapped onto essentially 
one point of $P$. The embedding of the remaining vertices of $T$ will be possible in a planar way if and only
if the paths $\pi_i$ can be decomposed into groups of exactly $B$ vertices, which is equivalent to the original 
$3$-partition instance. The following point set $P$ and simple polygon $S$ will do the work (see Figure \ref{PointsFig}):

\begin{figure}
\setlength{\unitlength}{1.00mm}
\centering
\vspace{-4.5cm}
\hspace*{-3cm}
\includegraphics{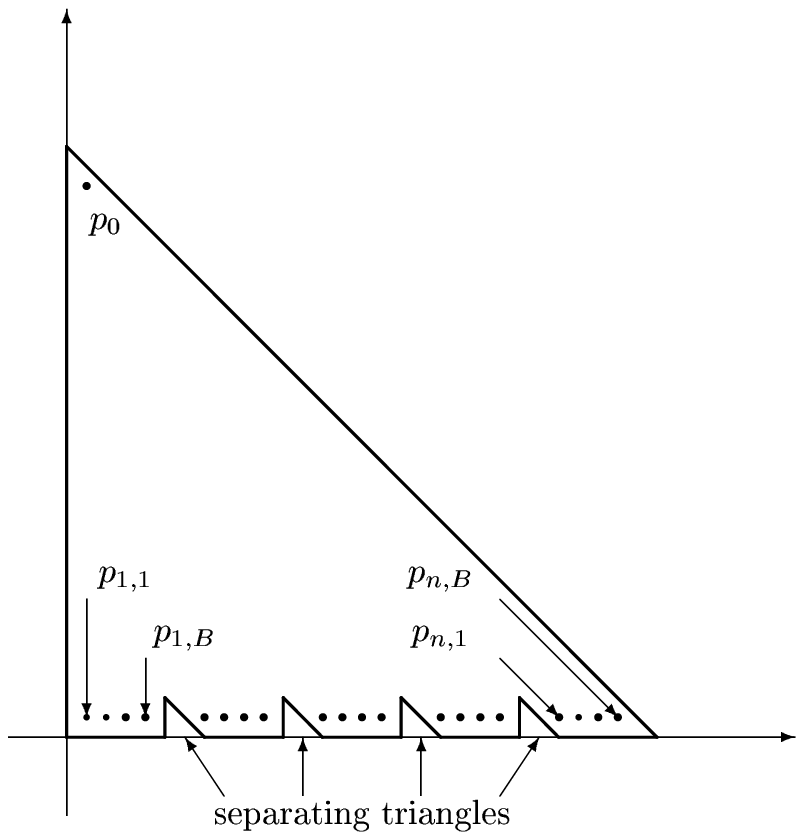}
\vspace{-17cm}
\caption{Point set P for the NP-hardness reduction.}
\label{PointsFig}
\end{figure}

\begin{itemize}
\item{Consider the simple polygon $S$ as a triangle constructed on three points $(0,0)$, $(n(B+2),0)$ and $(0,n(B+2))$, 
from which $n-1$ separating triangles are excluded. The coordinates of the three vertices of each separating triangle 
are $(B+1+k(B+2), 0)$, $(B+1+k(B+2), 2)$ and $(B+3+k(B+2), 0)$, where $0 \leq k \leq n-2$.}
\item{Consider $nB+1$ points inside simple polygon $S$ as follows. A single point $p_0$ with coordinates $(1, n(B+2)-2)$, 
in addition to $n$ groups of $B$ points. Point $p_{i,j}$, which is $j$th point of $i$th group, has coordinates 
$((i-1)(B+2)+j, 1)$, where $1 \leq j \leq B$ and $1 \leq i \leq n$.}
\end{itemize}

Simple polygon $S$ is constructed such that the points of each group are invisible from the points of other groups. 
The only point which is visible from the other points is point $p_0$. 

It is clear that in any straight-line planar embedding of $T$ onto the point set $P$ such that the edges of $T$ do not intersect 
the boundary of simple polygon $S$, then vertex $v_0$ has to be mapped onto point $p_0$. Otherwise, there would be one edge 
$(v_0, v_{i,1})$ that intersects the boundary of bounding simple polygon $S$.
Also, in such an embedding, the vertices of path $\pi_i$ of tree $T$ have to be mapped completely onto the points of 
just one group, otherwise the path intersects the boundary of $S$. Therefore, a geometric planar embedding 
of $T$ is possible if and only if paths of $T$ can be arranged in groups such that each group 
has exactly $B$ vertices. Recall that tree $T$ has $3n$ paths in addition to vertex $v_0$, for each 
path $\pi_i$ we have $|\pi_i| = a_i$, and point set $P$ consists of $n$ groups of $B$ points in addition 
to point $p_0$. Hence, any mapping of the vertices of paths of $T$, onto the points of $P$, 
provides a geometric planar graph inside simple polygon $S$ if and only if the $3$-partition 
instance has a solution. 

$3$-partition is NP-hard even when $B$ is bounded by a polynomial in $n$. Tree
$T$ has a polynomial number of vertices, and point set $P$ also has a polynomial number of
points. Furthermore, the coordinates of the points in $P$ are bounded by polynomials and the
whole reduction can be done in polynomial time. This finishes the proof of Theorem \ref{Theorem1}.

\section{Conclusion}
\label{ConclusionSect}
In this paper, we proved that given a free tree $T$ and a point set $P$ in the plane bounded by 
a simple polygon $S$, deciding whether there is a planar straight-line embedding of $T$ onto $P$ 
while the edges of $T$ do not intersect the boundary of $S$ is NP-complete. This problem is polynomially solvable 
\cite{B:OEOPG,BMS:OAET}, if there is not any bounding polygon.


\end{document}